\documentclass[a4paper,11pt]{article}

\usepackage{contribution}

\newcommand{\weblink}[2][]{%
    \ifthenelse{\equal{#1}{}}%
    {\textnormal{\url{#2}}}%
    {\textnormal{\href{#2}{#1}}}%
}

\newcommand{\acknowledgements}[1]{%
  \bigskip\bigskip
  \textsf{\textbf{\Large Acknowledgements}} \\[2ex]
  {#1}
  \bigskip
}

\def\beq{\begin{equation}}
\def\eeq#1{\label{#1}\end{equation}}
\def\eeqn{\end{equation}}

\def\beqa{\begin{eqnarray}}
\def\eeqa#1{\label{#1}\end{eqnarray}}
\def\eeqan{\end{eqnarray}}

\let\bar=\overbar

\def\tr{{\mbox{\rm tr}}}

\def\Dslash{\not{\hbox{\kern-4pt $D$}}}
\def\dslash{\not{\hbox{\kern-2pt $\del$}}}

\def\ee{e^+e^-}

\def\msb{{\bar{\ssstyle M \kern -1pt S}}}

\newcommand{\contribution}[7][]{%
  \clearpage
  \thispagestyle{plain}
  \ifthenelse{\equal{#1}{}}
  {\hypersetup{pdftitle={#2}}}
  {\hypersetup{pdftitle={#1}}}
  \hypersetup{pdfauthor={{#3} {#4}}}
  {\centering\normalfont\LARGE\bfseries\sffamily #2 \par\nobreak}
  \lhead{}
  \chead{%
    \textit{\footnotesize XIV International Conference on Hadron Spectroscopy
      (\weblink[\textit{hadron2011}]{http://www.hadron2011.de}), 13-17 June 2011, Munich, Germany}%
  }
  \rhead{}
  \bigskip
  \begin{center}
    {#3} {#4}\ifthenelse{\equal{#6}{}}{}{\footnote{\weblink[#6]{mailto:#6}}}
    \ifthenelse{\equal{#7}{}}{}{#7} \\
    \textit{#5}
  \end{center}
  \bigskip
}

\renewcommand{\abstract}[1]{%
  \begin{center}
    \begin{minipage}{0.85\textwidth}
      \begin{footnotesize}
        #1
      \end{footnotesize}
    \end{minipage}
  \end{center}
  \bigskip
}

\begin{document}

%
%
%
%
%
{  


\newcommand{\tb}{} 
\newcommand{\ts}{} 
\renewcommand{\tr}{} 
\newcommand{\nB}[1]{n_\rmi{B{#1}}}
\newcommand{\nF}[1]{n_\rmi{F{#1}}}
\newcommand{\rmO}{{\mathcal{O}}}
\newcommand{\mE}{m_\rmi{E}}
\newcommand{\fe}{\rmi{f}}
\newcommand{\bo}{\rmi{b}}
\newcommand{\eq}{eq.~}
\newcommand{\eqs}{eqs.~}
\newcommand{\fig}{fig.~}
\newcommand{\Tint}[1]{{\hbox{$\sum$}\!\!\!\!\!\!\int}_{\!\!\!\!#1}}
\renewcommand{\vec}[1]{{\bf #1}}
\newcommand{\nwc}{\newcommand}
\newcommand{\tinymsbar}{{\overline{\mbox{\tiny\rm{MS}}}}}
\nwc{\be}{\begin{equation}} 
\renewcommand{\ee}{\end{equation}} 
\nwc{\bmu} {\bar{\mu}}
\nwc{\ba}  {\begin{eqnarray*}}
\nwc{\ea}  {\end{eqnarray*}}
\nwc{\bi}  {\begin{itemize}}
\nwc{\ei}  {\end{itemize}}
\nwc{\nn}  {\nonumber\\}
\nwc{\Tr}  {\mathop{\rm Tr}}
\nwc{\re}  {\mathop{\rm Re}}
\nwc{\im}  {\mathop{\rm Im}}
\nwc{\Hc}  {\mathop{\rm H.c.}}
\nwc{\la}[1]{\label{#1}}
\nwc{\rmi}[1]{{\! \mbox{\scriptsize #1}}}
\nwc{\nr}[1]{(\ref{#1})}
\nwc{\fr}[2]{{\frac{#1}{#2}}}
\nwc{\msbar}{\overline{\mbox{\rm MS}}}
\nwc{\lambdamsbar}{\Lambda_{\overline{\rm MS}}}
\newcommand{\Nf}{N_{\rm f}}
\newcommand{\Nc}{N_{\rm c}}
\newcommand{\Tc}{T_{\rm c}}
\newcommand{\mention}[2]{\hfill\parbox[c]{#1}{\tiny \ts \hfill #2}}
\newcommand{\rmii}[1]{{\!\!\mbox{\tiny\rm{#1}}}}
\def\slash#1{#1\!\!\!/\!\,\,} 
\def\lsi{\raise0.3ex\hbox{$<$\kern-0.75em\raise-1.1ex\hbox{$\sim$}}}
\def\gsi{\raise0.3ex\hbox{$>$\kern-0.75em\raise-1.1ex\hbox{$\sim$}}}
\nwc{\lsim}{\mathop{\lsi}}
\nwc{\gsim}{\mathop{\gsi}}
\newcommand{\unit}{{\mathbbm{1}}} 

%

\contribution[News on hadrons in a hot medium]  
{News on hadrons in a hot medium}               
{M.}{Laine}                                  
{Faculty of Physics, University of Bielefeld, 
D-33501 Bielefeld, Germany}                     
{laine@physik.uni-bielefeld.de}                 
{}                                              
%

\abstract{%
 Some of the modifications that a thermal medium, of the type
 generated in heavy ion collision experiments at the LHC, may
 impose on the properties of hadrons, are reviewed. The focus 
 is on hadrons containing at least one heavy quark 
 (charm or bottom or their antiparticles).
}
%

\section{Introduction}

\vspace*{-9.5cm}

\hfill BI-TP 2011/19

\vspace*{9.0cm}

The title that I was given
is quite broad, so I chose to 
interpret it in a particular way. Denoting by $T$ the temperature
and by $\mu$ the baryonic chemical potential, the part 
``hot medium'' will be taken to indicate the temperature range
50 MeV $\ll T \ll$ 1000 MeV, perhaps reachable in the current generation
of heavy ion collision  experiments; the medium
is also assumed {\em not} to be ``dense'', meaning that the chemical
potential is small, $|\mu| \ll \pi T$. The most drastic
cut concerns ``hadrons'': in this talk only those which more or less
maintain their identity in the temperature range considered are discussed, 
meaning that at least one heavy quark with a mass $M \gg \pi T$ should be 
present. The final cut is related to the word ``news'':
I will try to concentrate on recent works from 2010/2011, 
not tracing literature any further back than to 2004 (with one exception).

With these excuses, the outline is that I will start by reviewing 
``open $c,b$'', i.e.\ the fate of $D$ and $B$ mesons in a hot medium. 
Subsequently recent developments concerning quarkonium physics, 
or ``bound $c\bar{c}, b\bar{b}$'', i.e.\ $J/\Psi$ and $\Upsilon$ mesons, 
are discussed. A final short section concerns what I refer to 
as ``thermal $c\bar{c}, b\bar{b}$'', standing for quark-antiquark pairs
generated from thermal fluctuations. I would like to apologize once more
for the exclusion of lighter hadrons, 
on which many new insights have been reported at this 
conference and elsewhere and but on which I possess no expertise.

%
\section{Open $c,b$}

Jets forming around energetic single heavy quarks are among the
basic events in a hadronic collision, and the gluon-fusion
amplitude that is responsible for them, illustrated at leading order as  
\be
 \begin{minipage}[c]{3.5cm}
 \epsfxsize=3.5cm\epsfbox{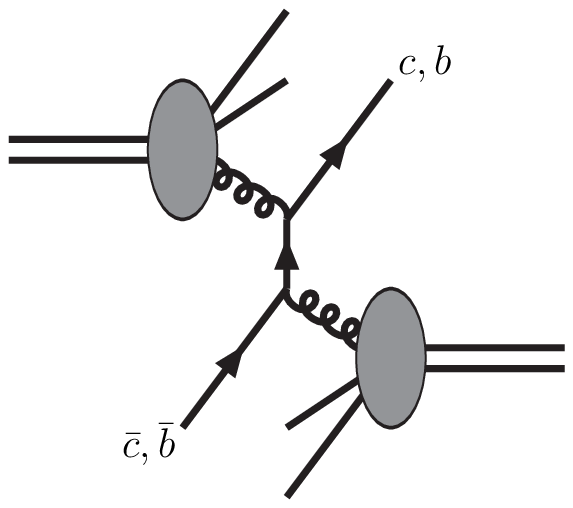} 
 \end{minipage} 
 \la{initial_open}
\ee
is theoretically relatively well understood~\cite{initial}. 
For thermal physics, the interesting 
question is what happens after the initial production. 
The idea is that apart from the heavy quarks, a lot of ``soft stuff'' gets 
generated as well, 
which could rapidly reach a state of 
thermal equilibrium. Then the heavy quarks need to propagate through 
this ``medium'' (shown with yellow below), 
a process that can be sketched as
\be
 \begin{minipage}[c]{6.5cm}
 \epsfxsize=6.0cm\epsfbox{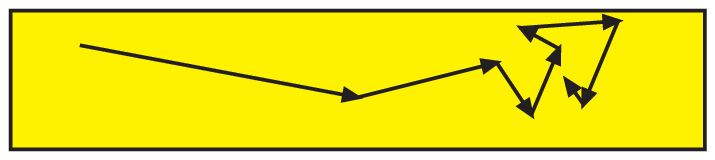} 
 \end{minipage} \;.
\ee
What is illustrated here is that 
the heavy quark jets tend to get slowed down and eventually 
stopped, by bremsstrahlung as well as by elastic scatterings. 
In the latter case some gluons can be off-shell and soft, and 
this can lead to large infrared effects from processes like  
\be
 \begin{minipage}[c]{1.5cm}
 \epsfysize=1.5cm\epsfbox{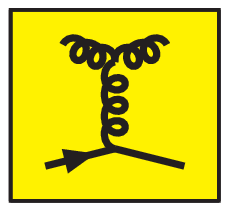} 
 \end{minipage} \quad \;.
\ee
Therefore we may expect, with some hindsight already folded in, 
that despite their large inertia the heavy quarks do interact 
strongly with the medium. In fact, experimentally, 
charm quark jets get ``quenched'' practically as effectively
as light jets;  this is shown in \fig\ref{exp:ALICE}, and 
is among the reasons that the medium generated 
in heavy ion collisions is nowadays conceived to constitute a 
``strongly interacting quark-gluon plasma'', or sQGP.

%
\begin{figure}[htb]
 
\vspace*{-0.2cm}

 \centerline{%
 \epsfxsize=7.0cm\epsfbox{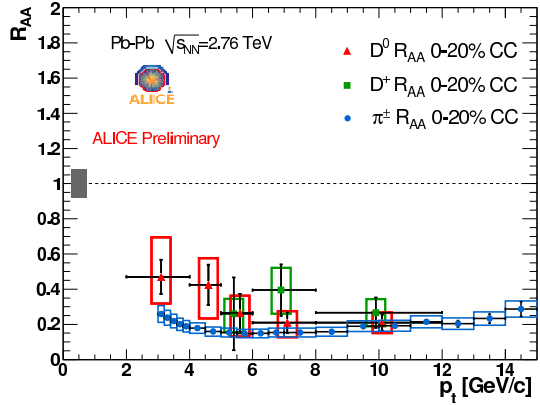} 
}
 \caption{The ``nuclear modification factor'', $R_{AA}$, as a function 
 of the transverse momentum, from recent ALICE data~\cite{Dainese:2011vb}.
 The fact that $R_{AA}<1$ indicates that jets get quenched, and the 
 data show that $D$-mesons stop about as effectively as pions despite
 their much larger inertia. In fact even $B$-mesons appear to interact
 strongly with the medium~\cite{Dainese:2011vb}.}
 \la{exp:ALICE} 

\end{figure}
%

It is a challenge for the theoretical finite-temperature
community to quantitatively explain, 
starting from the basic rules of QCD and statistical 
physics, the experimentally observed strong interactions felt by
the heavy quarks. (Although inspiring, 
AdS/CFT-based studies for analogous theories
are not discussed here for lack of space.)
Indeed many different lines of research have 
been pursued in this vain. For instance, 
a parameter that characterizes the strength of interactions felt 
by the heavy quarks, called the ``momentum diffusion coefficient'', 
has been computed at leading~\cite{mt} and next-to-leading~\cite{chm1,chm2}
order in the weak-coupling expansion. Given that the weak-coupling 
expansion shows questionable convergence, various model studies 
at $T \gg 200$ MeV also appear well-motivated as an intermediate
stage~\cite{vH,mink,rr}. In the end, though, it is important 
to come up with a non-perturbative first-principles formulation
within QCD~\cite{ct,eucl} and move towards a numerical lattice 
determination of the quantities in question~\cite{rhoE,hbm,analytic}.
(For charm this is also being attempted 
through a ``brute force'' approach~\cite{ding2}.)
Another line is that at low temperatures, $T \ll 200$ MeV, the same
physics can be addressed analytically through the use of chiral 
effective theories or extensions thereof~\cite{hadronic,had1,had2,had3}.
There are also many works in continuous progress, with a basic philosophy 
that can be traced back to ref.~\cite{mt}, that aim to implement theoretical 
results of the type mentioned 
in a combined hydrodynamics and Langevin simulation, 
in order to obtain results that can be directly
compared with experiment.

Let me highlight a couple of these developments in a bit more detail. 
A perhaps most economical non-perturbative
formulation of the problem, valid at least for the bottom case, is to make use 
of Heavy Quark Effective Theory in order to remove the heavy quark mass
scale from the problem, leaving over a purely gluonic correlator~\cite{eucl}: 
\be
 G_E(\tau) \equiv - \fr13 \sum_{i=1}^3 
 \frac{
  \Bigl\langle
   \re\Tr \Bigl[
      U_{\beta;\tau} \, gE_i(\tau,\vec{0}) \, U_{\tau;0} \, gE_i(0,\vec{0})
   \Bigr] 
  \Bigr\rangle
 }{
 \Bigl\langle
   \re\Tr [U_{\beta;0}] 
 \Bigr\rangle
 }
 \;, \la{GE_final}
\ee
where $gE_i$ stands for the (bare) 
colour-electric field and $U_{\tau_2;\tau_1}$
is a straight timelike Wilson line in Euclidean signature. The ``transport
coefficient'' corresponding to this correlator yields the momentum 
diffusion coefficient alluded to above, often denoted by $\kappa$.
In \fig\ref{fig:kappa} results are shown from the direct next-to-leading
order perturbative computation of $\kappa$ and from a measurement of 
the correlator defined in \eq\nr{GE_final}. It should become clear 
how large corrections (left) could be pretty much hidden
on the Euclidean lattice (right), whose results at first sight follow
closely a next-to-leading order prediction. This indicates
that a very high resolution is needed in the lattice measurement; 
therefore no final results are available to date, but the problem
appears (relatively) well-posed, and work continues. 

\begin{figure}[htb]

 \begin{minipage}[c]{7cm}
 \includegraphics[width=7cm,height=6.0cm]{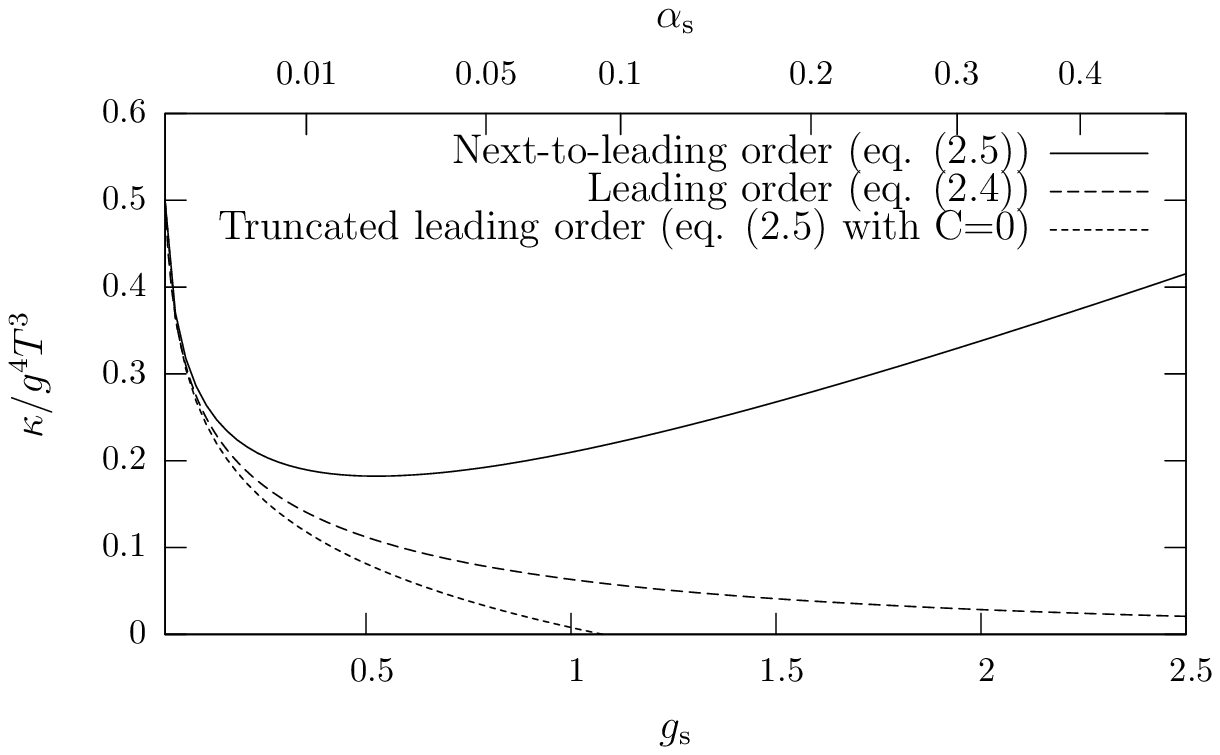}
 \end{minipage}~\hspace*{5mm}~
 \begin{minipage}[c]{7.0cm}
 \epsfxsize=7.0cm\epsfbox{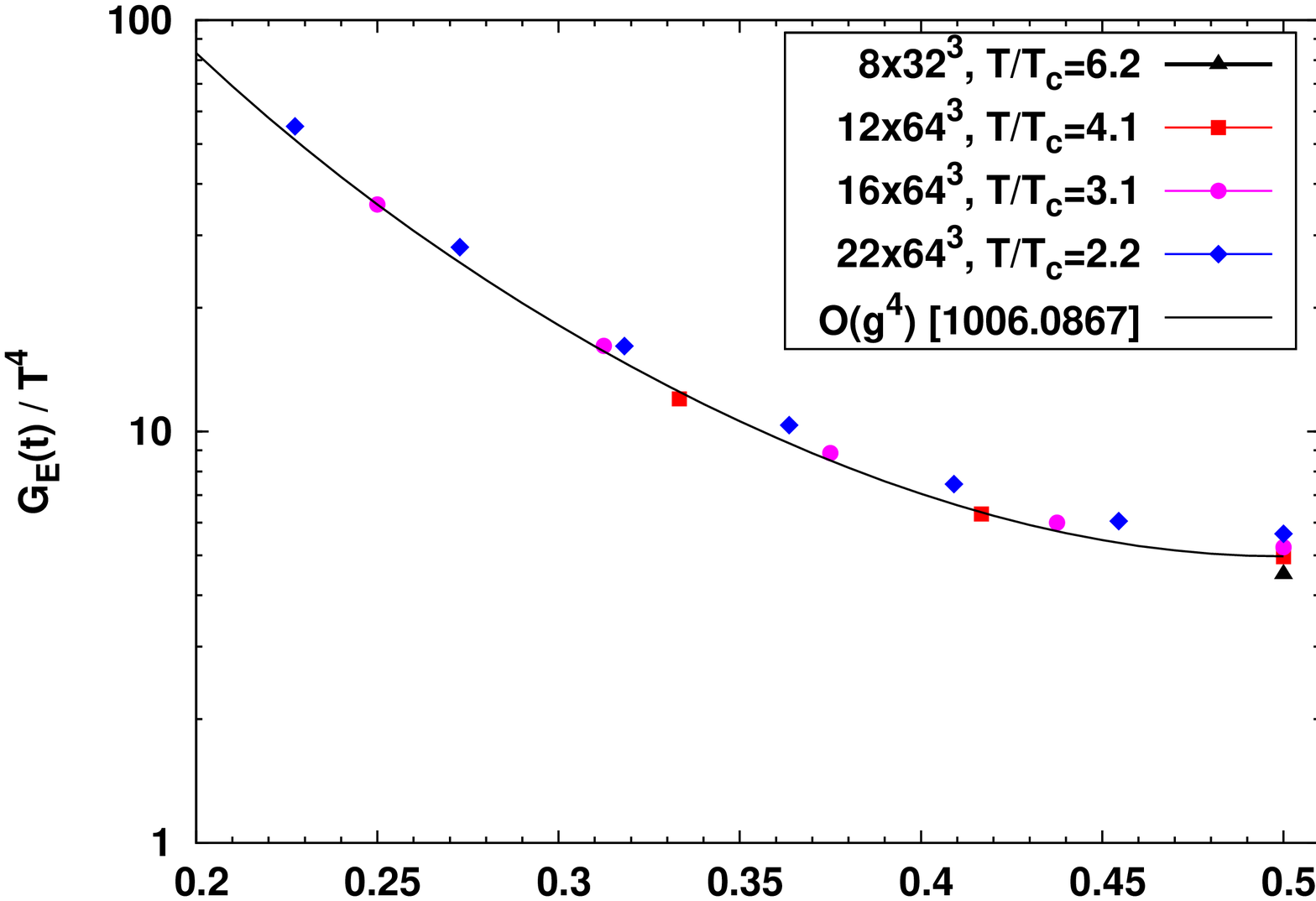} 
 \end{minipage}
 \caption{Left: Leading and next-to-leading order results for the heavy-quark
 momentum diffusion coefficient, $\kappa$, as a function of 
 a renormalized gauge coupling (from ref.~\cite{chm2}). 
 Right: Lattice results for the corresponding Euclidean
 correlator (from ref.~\cite{hbm}), compared 
 with a next-to-leading order determination of the same quantity
 (from ref.~\cite{rhoE}).}
 \la{fig:kappa}
\end{figure}
%

As another highlight, consider the hadronic phase, $T \ll 200$ MeV. 
The momentum diffusion coefficient mentioned above can be related
(in the limit $M \gg \pi T$) to the ``usual'' diffusion coefficient, let us 
denote it by $D$, through the Einstein relation, $D = 2 T^2/\kappa$.
If interactions are strong, then $\kappa$ is large 
(cf.\ \fig\ref{fig:kappa} left)
and therefore $D$ is small. In \fig\ref{fig:comparison}, results are 
shown from various works, concentrating particularly on 
the hadronic phase; the results have been obtained by making
use of Heavy Meson Chiral Perturbation Theory, hadronic
models, or some ``intermediate'' framework. The main observations are that, 
first of all, various works differ signficantly from each other so that
they probably contain non-negligible systematic uncertainties; but that, 
second, it seems conceivable that strong interactions are {\em not}
immediately switched off in the hadronic phase. This might play a role
in the analysis of heavy ion collision data as well.  

%
\begin{figure}[htb]

\centerline{
 \epsfxsize=8.0cm\epsfbox{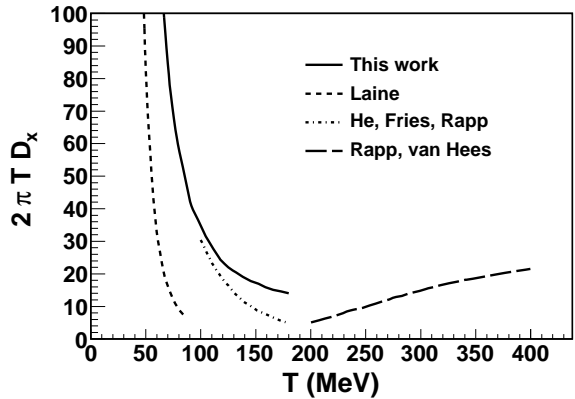} 
}

 \caption{The charm quark diffusion 
 coefficient from ref.~\cite{had3}, with results 
 from refs.~\cite{hadronic,had1,vH} also shown for comparison.} 
 \la{fig:comparison} 

\end{figure}
%

%
\section{Bound $c\bar{c}, b\bar{b}$}

Although quarkonium physics is in some sense a more traditional
and popular probe for quark-gluon plasma formation than the physics related
to heavy quark jets discussed in the previous section, it is also 
clear that this is a more challenging topic, both theoretically
and experimentally. The difficulty becomes apparent already by drawing
a simple example for an initial hard process in which quarkonium 
can get generated: 
\be
 \begin{minipage}[c]{3.5cm}
 \epsfxsize=3.5cm\epsfbox{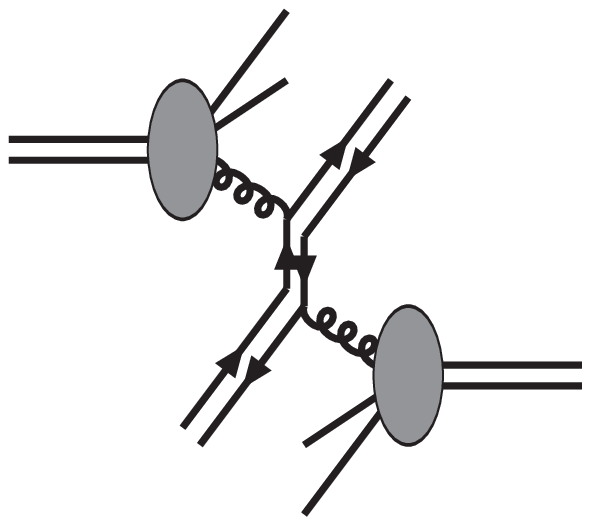} 
 \end{minipage}
 \la{initial_bound} \quad \;.
\ee
Clearly the cross section must be smaller than that in \eq\nr{initial_open}, 
and furthermore the corresponding rate is more difficult to compute reliably. 
On a more positive note, the final signal is simple, given that
quarkonium can decay into a dilepton pair, say $\mu^+\mu^-$, which is easily
identified and whose invariant mass can be measured with a relatively
good resolution: 
\be
 \begin{minipage}[c]{5.5cm}
 \epsfxsize=5.5cm\epsfbox{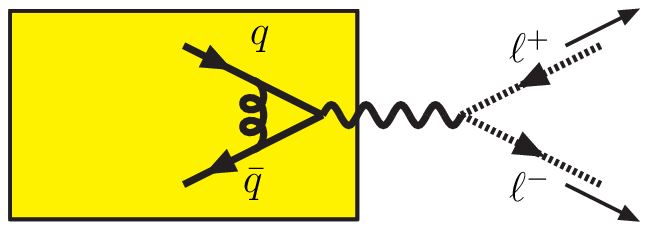}~~~%
 \end{minipage} \quad \;.
\ee
Indeed a possible thermal modification 
of the bottomonium spectral shape, 
illustrated in \fig\ref{fig:CMS}, is among the  
most spectacular early results from the LHC heavy ion program.

%
\begin{figure}[htb]

 \vspace*{3cm}

 \epsfxsize=4.1cm\epsfbox{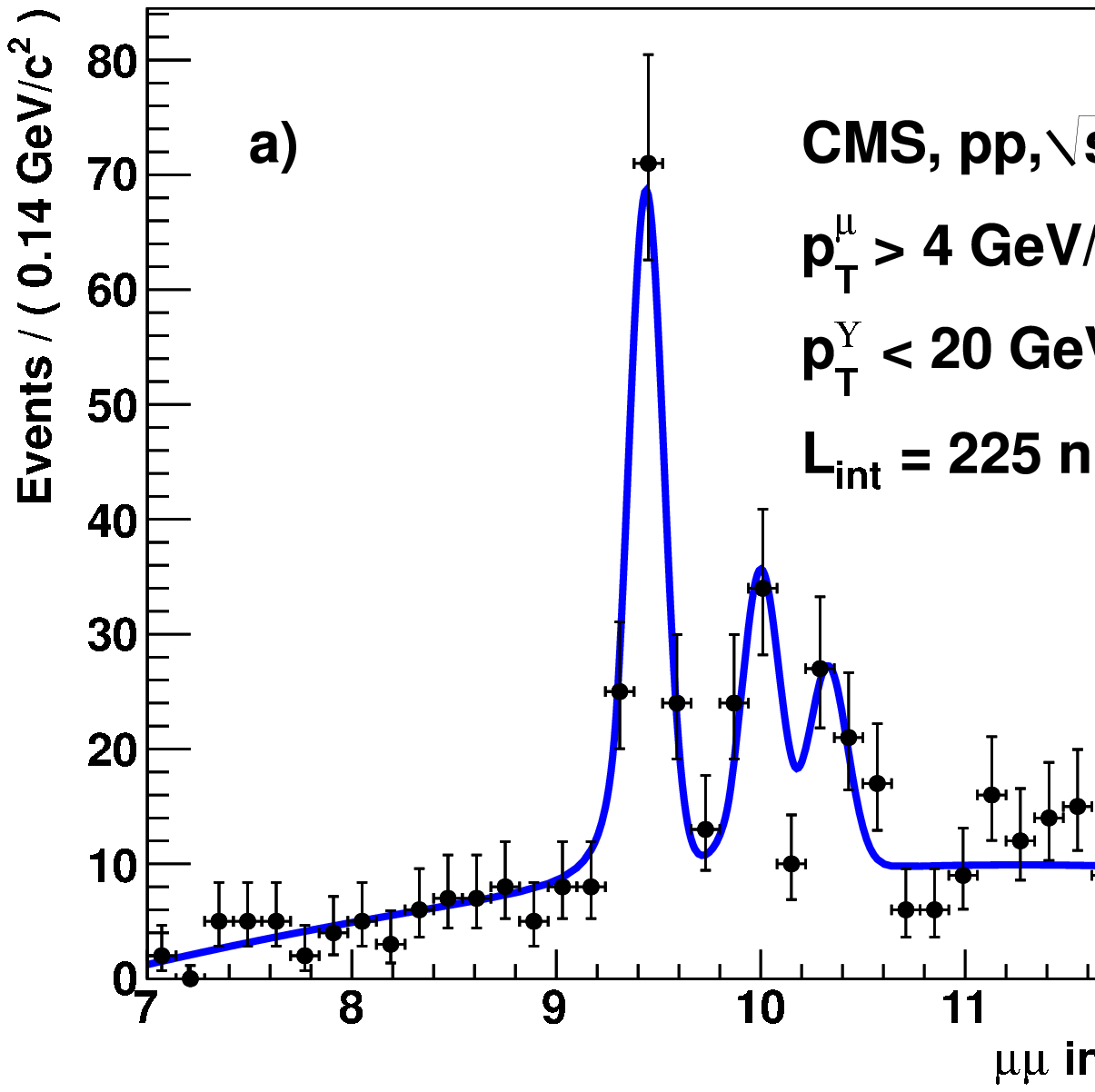}%
 \hspace*{3.5cm}
 \epsfxsize=4.1cm\epsfbox{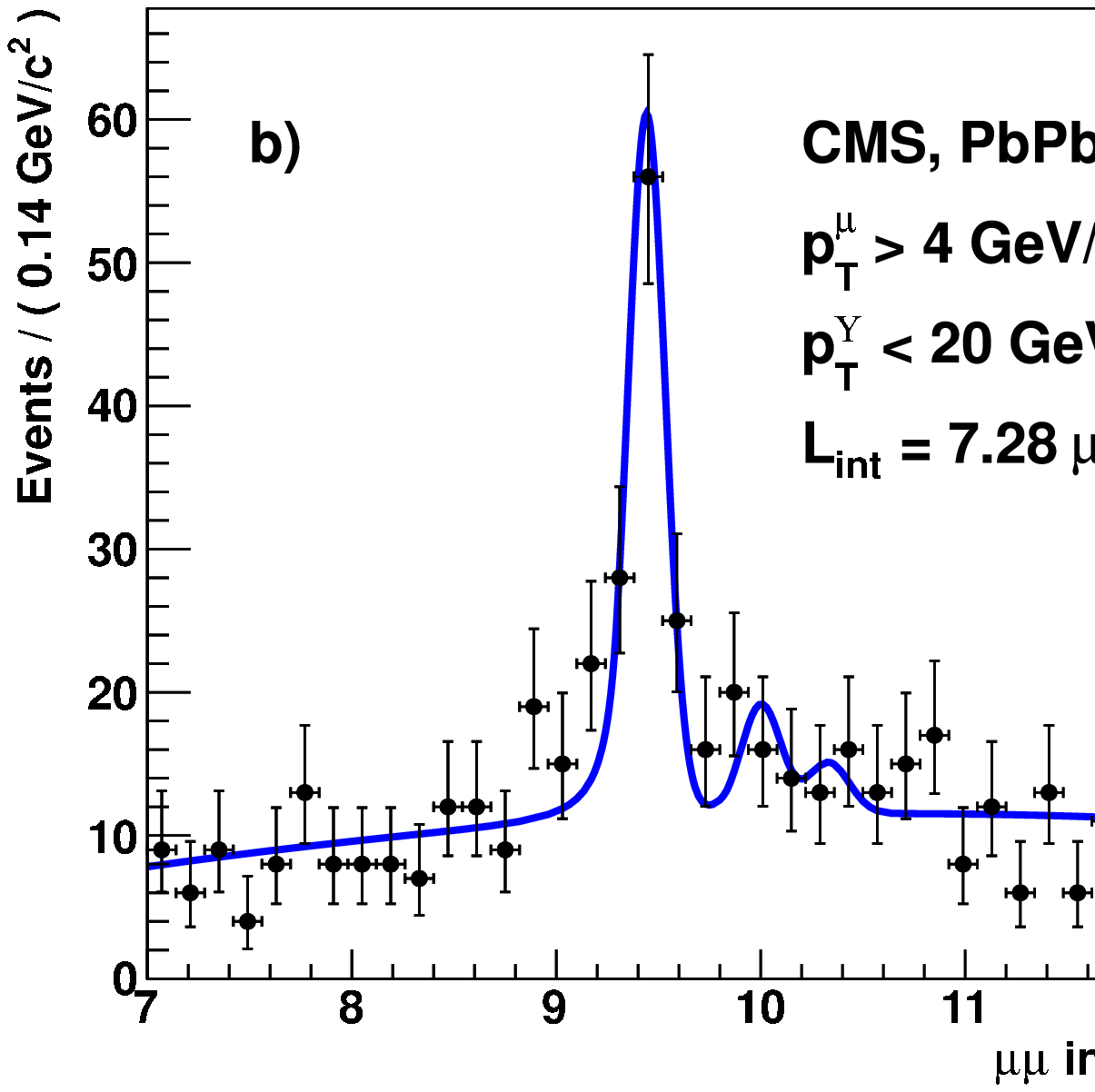}%
 
 \vspace*{1cm}

 \caption{The disappearance of spectral features from  
 bottomonium resonances when going from {\em pp} to {\em PbPb} collisions, 
 according to CMS (from ref.~\cite{Chatrchyan:2011pe}).
 Qualitatively the pattern appears to be in accordance with 
 a classic ``sequential suppression'' scenario~\cite{kks} 
 (see also ref.~\cite{nb}); 
 a quantitative interpretation has been put forward in ref.~\cite{ms}. }
 \la{fig:CMS}

\end{figure}
%

Apart from experimental news, there has also been some
theoretical progress on thermal quarkonium physics
in recent years.
I would like to imagine that a slight paradigm shift is taking 
place: whereas it was originally 
envisaged that quarkonium remains a coherent 
quantum-mechanical bound state in a thermal medium, with only 
the potential that binds it together getting modified by 
Debye screening~\cite{masa}, a process that could perhaps 
be illustrated as \vspace*{-5mm}
\be
 \begin{minipage}[c]{6.5cm}
 \hspace*{0.5cm}
 \epsfxsize=5.5cm\epsfbox{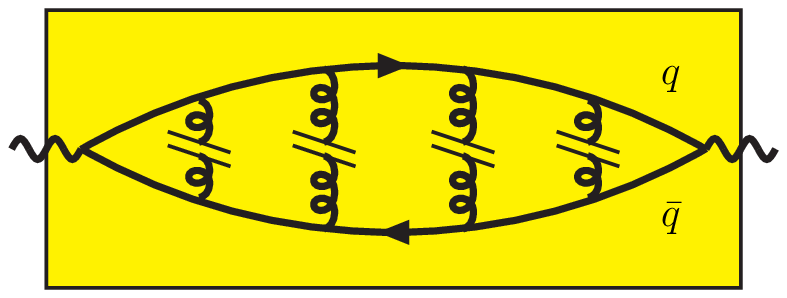} 
 \end{minipage} \quad \;, \la{trad}
\ee
a ``modern'' view is that there might be an additional effect 
also taking place: coherence may be (partly) lost 
due to random kicks from a heat bath. Technically, this implies
that a suitably defined static potential may develop an imaginary part. 
The corresponding physics could be thought of for instance
in terms of the following diagram, obtained from \eq\nr{trad}
by twisting two of the gluon lines outwards: 
\be
 \begin{minipage}[c]{4.0cm}
 \epsfxsize=4.0cm\epsfbox{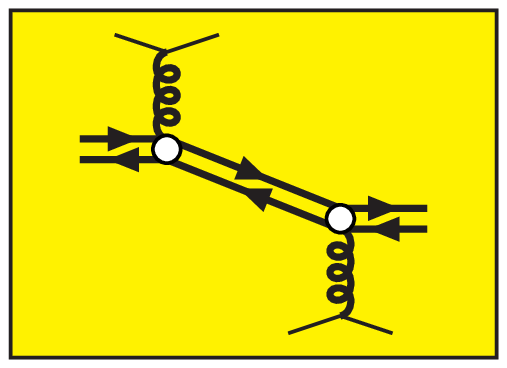}~~~%
 \end{minipage} \quad \;.
\ee

The full body of literature concerning quarkonium
at high temperatures is too vast to fit in this review, so I restrict to 
mentioning representative theoretical works that contain 
elements in this ``modern'' direction. Various computations resulting
in a complex real-time static potential have been reported e.g.\ in refs.~%
\cite{static,imV,beraudo,akt,noneq,dumitru1,%
dumitru2,tassler,blaizot,chandra,margotta}
(actually many inequivalent definitions of 
a ``static potential'' are being used).
These results can be given a systematic role at least in the
weak-coupling regime where a number of momentum scales can 
be identified and, if there is a hierarchy between them, 
a full-fledged effective field theory approach can be 
formulated (termed ``PNRQCD$_\rmi{HTL}$'' in a particular regime)~%
\cite{es1,nb1,es2,nb2}.
On the side of applications, at high temperatures where quarkonium
``melts'' the quantities that can still be defined are the 
spectral function 
and the thermal part of the dilepton production rate, 
${\rm d}N_{\mu^-\mu^+} / {\rm d}^4 x\, {\rm d}^4 Q$~%
\cite{kl,pert,peskin,sewm08,nlo,grigo,rr,miao}.
On the other hand, if we stay below the melting temperature, 
which may be phenomenologically relevant particularly for some of the 
bottomonium resonances (cf.\ \fig\ref{fig:CMS}), 
then it is meaningful to speak of the ``binding
energy'' of a bound state as well as of its ``width''; these are
simpler objects than the complete spectral function, and can be 
computed in rather explicit form~\cite{nb3}, including even 
their velocity dependence~\cite{es3} (see also ref.~\cite{dw}). 
Ultimately, of course, 
a lattice investigation is needed~\cite{jakovac,aa1,ding1,ohno}, 
but even those might preferably be formulated within 
a non-relativistic effective theory approach~\cite{rothkopf,ar2,aa2}, 
particularly in the bottomonium case. 

%
\begin{figure}[tb]

 \hspace*{5mm}
 \begin{minipage}[c]{7.0cm}
 \epsfysize=5.8cm\epsfbox{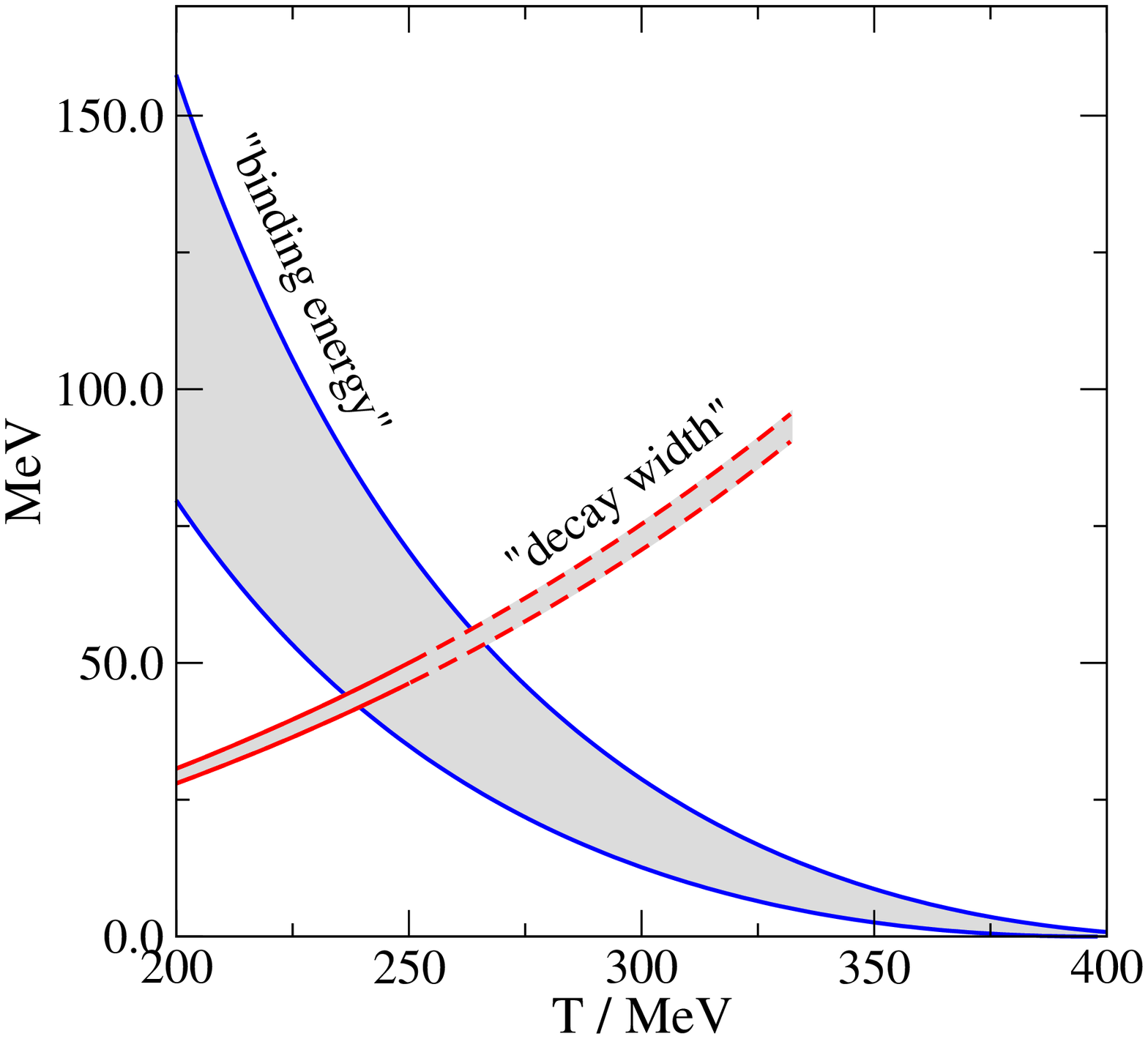}%
 \end{minipage}%
 \hspace*{5mm}
 \raise1ex%
 \hbox{\begin{minipage}[c]{7.0cm}
 \epsfysize=6.3cm\epsfbox{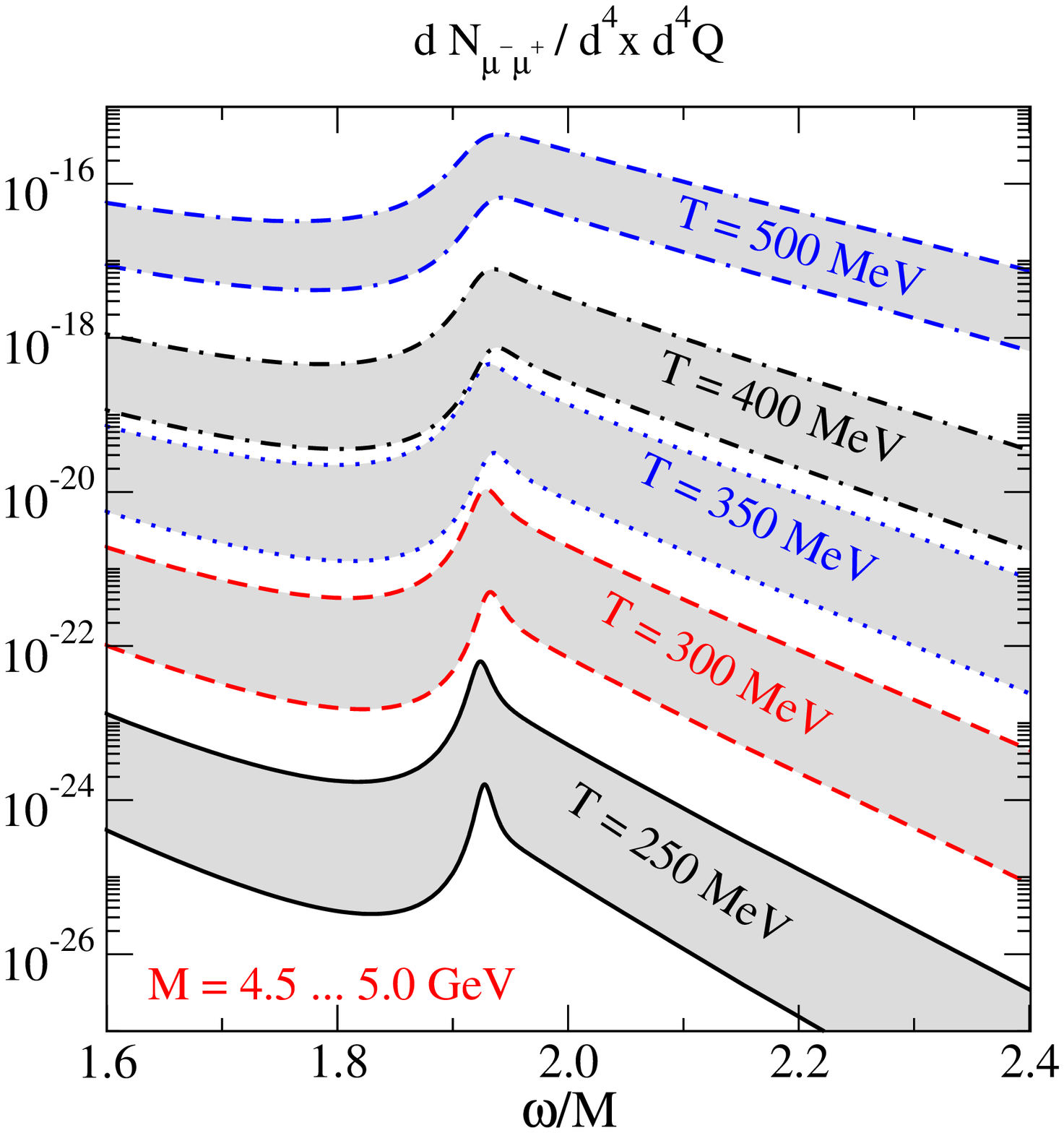}%
 \end{minipage}}
 
 \vspace*{5mm}

 \caption{Left: A thermal ``decay width'' that decoherence due to 
 random kicks from the heat bath could bestow upon a quantum-mechanical
 bound state (from ref.~\cite{static}). Right: A thermal component 
 in the dilepton production rate, with a resonance peak at low 
 temperatures going over into a smooth threshold at high 
 temperatures (from ref.~\cite{nlo}).}
 \la{fig:width}

\end{figure}
%

Let me again highlight a couple of these developments in a bit more 
detail. The conceptual change from an ``on-off'' melting picture 
towards a gradual evolution of the spectral shape of a quarkonium
resonance is illustrated on the qualitative level in~\fig\ref{fig:width}, 
for the bottomonium case. More quantitatively, at low temperatures, 
far below ``melting'' ($T \ll 250$~MeV in terms of a situation
illustrated in \fig\ref{fig:width}(left)), the width shows  
a specific $T$-dependence~\cite{nb3} and, as mentioned, 
its velocity-dependence is also computable~\cite{es3}.

As a second highlight, let me mention lattice computations 
within effective theories. For lattice QCD, a ``scale hierarchy'' 
(say, between $m_\pi$ and $m_B$)
is always a serious challenge, because for a reliable representation 
of continuum physics the lattice spacing should be shorter
than the smallest physical length scale of the system, whereas for
an exclusion of finite-volume effects the lattice extent should
exceed the largest physical length scale; these complementary
requirements easily lead to a prohibitively large 
number of lattice sites. 
On the other
hand, for effective field theories, the existence of a clear scale 
hierarchy is a blessing. This suggests the idea of combining the 
two approaches; indeed a study of the bottomonium system within 
the so-called NRQCD approach has recently been launched~\cite{aa2}
(there are well-known issues related to the existence
of a continuum limit but these should be no worse than at zero temperature). 
Another direction is to directly determine
a real-time static potential, perhaps to be used within
a ``PNRQCD$_\rmi{HTL}$''-type framework, 
through a spectral analysis 
of an imaginary-time Wilson loop~\cite{rothkopf,ar2}, as sketched 
in \fig\ref{fig:imV}.

%
\begin{figure}[tb]

 \hspace*{4cm}%
 \begin{minipage}[c]{10.0cm}
 \epsfysize=4.0cm\epsfbox{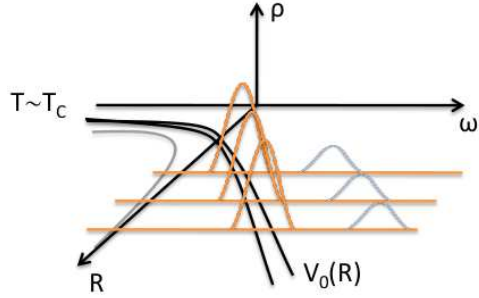} 
 \end{minipage}
 \caption{
   A sketch for how a spectral analysis of an imaginary-time Wilson 
   loop could reveal a real part
   of a static potential, from the position of a peak, 
   and an imaginary part, from its width
   (from a talk by A.~Rothkopf, 
   illustrating the work reported in refs.~\cite{rothkopf,ar2}).
  }
  \la{fig:imV}

\end{figure}
%

%
\section{$c\bar{c},b\bar{b}$ from thermal fluctuations}

It is interesting to ask under which conditions
heavy quarks or mesons can ``chemically equilibrate'',
i.e.\ be part of the heat bath with an entropy density
determined by $T$, rather than by some 
non-thermal initial process such as those illustrated 
in \eqs\nr{initial_open}, \nr{initial_bound}. Naively, 
one would think that this is the case only for 
$T \gg 2 M$, so that there is no Boltzmann suppression, 
$\exp(-\frac{2M}{T})$, hindering the rate of 
pair creation of a quark-antiquark pair.
However, trying to be more quantitative, one may ask e.g.\ how 
the heavy quark mass $M$ should be interpreted; say, as 
an $\msbar$ scheme mass, $M\to m_c^{\raise-0.2em\hbox{$\tinymsbar$}}
(\mbox{3~GeV}) \approx 1$~GeV; as a more ``physical'' 
pole mass,  
$M\to m_c^\rmi{pole} \sim (1.5-2.0)~\mbox{GeV}$; 
or as something else? Furthermore, the Boltzmann weight
comes with a prefactor which might partially compensate
for the exponential suppression, and in fact anyone familiar
with the ``imaginary-time'' formalism of thermal field theory
would from the outset suggest a comparison of $2M$ with 
$2\pi T$ rather than $T$, which is a significant difference. 
All in all, it seems that the issue is non-trivial.

%
\begin{figure}[htb]

\hspace*{5mm}%
\begin{minipage}[c]{6.5cm}
 
 \epsfxsize=6.0cm\epsfbox{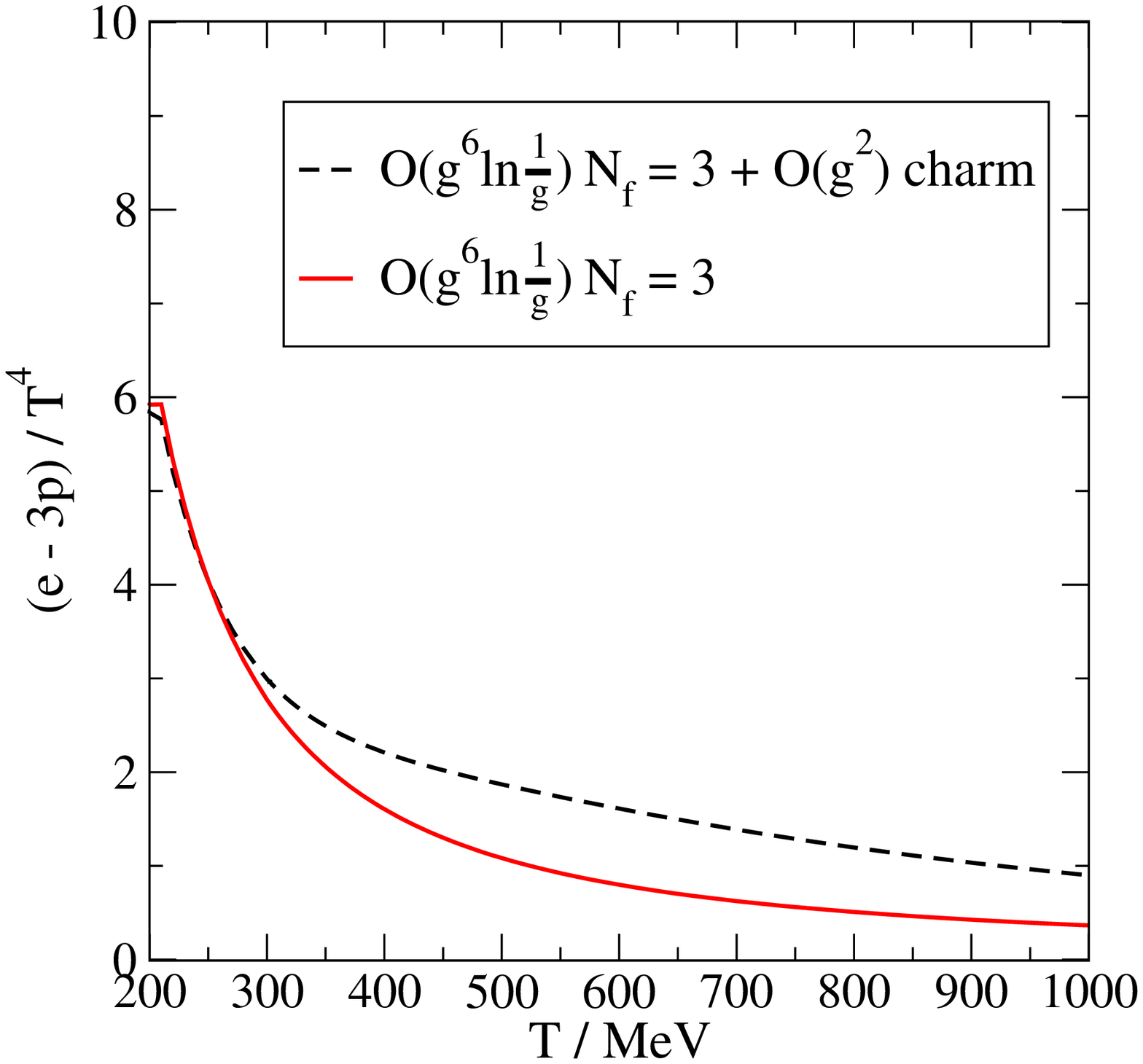} 
 
\end{minipage}~~~%
\begin{minipage}[c]{7.5cm}


 \vspace*{3mm}
 \hspace*{-3mm}\epsfxsize=7.5cm\epsfbox{EOS_charm} 

 \end{minipage}%

 \caption{Left: Perturbative results for the ``trace anomaly'',
 with and without charm quarks 
 (from ref.~\cite{phenEOS}). Right: Lattice estimates
 for the same quantity, denoted here by $I$ 
 (from ref.~\cite{lat1}; see  
 refs.~\cite{lat2,lat3} for similar works by other groups).}
 \la{fig:phenEOS}
  
\end{figure}
%

In any case, 
sample results from computations of how chemically equilibrated 
charm quarks would contribute to thermodynamic observables
are shown in \fig\ref{fig:phenEOS}, both from the weak-coupling
expansion and from the lattice. It seems that there
could be effects visible at surprisingly low temperatures; 
this might be relevant for the initial stages of 
hydrodynamics at the LHC, where current phenomenological 
studies tend to ignore charm quarks altogether. (For another 
argument in a related direction, see ref.~\cite{gt}.)

%
\section{Conclusions}

In heavy ion collisions at the LHC, heavy-quark 
related observables are becoming increasingly important: 
the existence of the heavy mass 
scale makes experimental signals clearly identifiable, and
also facilitates theoretical analyses by allowing for the use of 
modern effective theory methods. 
That said, much further work is needed for quantitative conclusions.

%
\acknowledgements{%
 This work was partly supported by the BMBF under project
 {\em Heavy Quarks as a Bridge between
      Heavy Ion Collisions and QCD}.
}


%

}  


\end{document}